\documentclass[journal]{IEEEtran}
\usepackage{courier}
\usepackage{soul}
\usepackage{color,url,xspace}
\usepackage{amssymb,amsmath}
\usepackage{graphicx}
\usepackage{colordvi}
\usepackage{epsfig}
\usepackage{endnotes}
\usepackage{verbatim}
\usepackage{subfigure}
\usepackage{epstopdf}
\usepackage{textcomp}
\usepackage {subfig}
\usepackage{setspace}
\usepackage{color}
\usepackage{soul}

\begin{document}
\title{Max-Weight Scheduling and Quality-Aware Streaming for Device-to-Device Video Delivery}

\author{Joongheon~Kim,~\IEEEmembership{Member,~IEEE, }
	Andreas~F.~Molisch,~\IEEEmembership{Fellow,~IEEE, }
	Giuseppe~Caire,~\IEEEmembership{Fellow,~IEEE}
\thanks{J. Kim is with the Department of Computer Science, University of Southern California, Los Angeles, CA 90089, USA e-mail: joonghek@usc.edu.}
\thanks{G. Caire and A.F. Molisch are with the Department of Electrical Engineering, University of Southern California, Los Angeles, CA 90089, USA e-mails: \{caire, molisch\}@usc.edu.}
}
\maketitle

\begin{abstract}
We propose and analyze centralized and distributed algorithms for device-to-device video scheduling and streaming.
The proposed algorithms address jointly
the problems of device-to-device link scheduling and
video quality adaptation in streaming.
Our simulations show that the proposed algorithms significantly outperform conventional separated approaches that treat these two problems independently.
\end{abstract}

\section{Introduction}
Due to the increased video traffic, efficient video-aware transmission schemes are of highest importance~\cite{cm2013golrezaei}.
In particular, video throughput of wireless networks can be enhanced by device-to-device (D2D) communications~\cite{jsac2014ji}.

D2D video transmission needs to deal with two aspects: (i) link scheduling and (ii) adaptive video quality in streaming. Traditionally these aspects have been treated separately. In this paper we propose and analyze algorithms that take the interrelationship between those tasks into account.

For \textit{centralized} scheduling, the scheduling decisions can be made by a base station, assuming it knows the channel states for all the possible D2D pairs.
We can formulate the scheduling problem as a max-weight independent set (MWIS) problem which we solve (approximately) using a well-established message-passing algorithm~\cite{tit2009sanghavi}.
The weights for the MWIS problem arise from the streaming (quality control) part.

Currently, the most well-known \textit{distributed} D2D scheduler is \textit{FlashLinQ}~\cite{ton2013wu}.
However, FlashLinQ does not incorporate naturally a video-quality-aware mechanism, and therefore its suitability for D2D \typeout{on-demand }video streaming remains open.
We propose an improvement by incorporating weights.

Our \textit{streaming} algorithm dynamically controls the quality mode of each chunk to maximize the total quality subject to
    all data being supportable. It does so via a stochastic formulation that can trade off queue stability with delay.

Obviously, higher video quality requires transmission of more bits, which  increases the probability of playback stalls. The tradeoff between video quality and stall probability is a key measure for any transmission scheme, and we use it to assess the performance of the algorithms in this paper. Our simulations show that algorithms that exploit the interconnection between scheduling and streaming provide a significantly better performance, i.e., better tradeoff between stall probability and video quality.



\section{The Proposed Algorithms}
Each D2D TX has a queue whose length evolves as:
\begin{equation}
Q_{i}(t+1) = \max\left[0, Q_{i}(t) - \mu_{i}(t)\right] + \lambda_{i}(t), l_{i}\in\mathcal{L}
\label{eq:q}
\end{equation}
where $Q_{i}(t)$, $\mu_{i}(t)$, and $\lambda_{i}(t)$ stand for
    the queue backlog size at the TX of the $i$-th D2D link $l_{i}$ at time $t$,
    the number of bits leaving the queue of the TX of $l_{i}$,
    and the number of bits added to the queue of the TX of $l_{i}$ (associated with the placement of chunks), respectively.

Suppose that a chunk (the basic transmission unit, about 0.5 sec. of video) contains $\mathcal{N}$ pixels and each chunk of each file $f$ is encoded at a number of different quality modes $q \in M=\left\{q_{1}\cdots q_{M}\right\}$.
Let $\mathbb{P}_{f}(q, \tau)$ and $\mathcal{N}\mathbb{B}_{f}(q, \tau)$ denote
    the video quality measure (e.g., peak-signal-to-noise-ratio (PSNR))
    and the number of bits for $f$ at chunk time $\tau$ with $q$, respectively.
A network controller
    chooses $q_{i}(\tau)$ for each $\tau$ for all requesting RXs $i$,
    allocates the source coding rate $\mathbb{B}_{f_{i}}(q_{i}(\tau), \tau)$.
The TX of $l_{i}$ places the $\mathcal{N}\mathbb{B}_{f_{i}}(q_{i}(\tau), \tau)$ bits in its TX queue $Q_{i}(\tau)$.

 \vspace{-1.0mm}

\subsection{Max-Weight Scheduling}

\textbf{Centralized scheduling with MWIS formulation:}
The scheduling is based on a link conflict graph such that the links scheduled to be active simultaneously in any time slot must form an independent set. In the conflict graph where the D2D links constitute the nodes, i.e., $l_{i}\in\mathcal{L}, \forall i\in\{1,\cdots,|\mathcal{L}|\}$ where $\mathcal{L}$ stands for the set of D2D links.
Thus, edge in the conflict graph between $l_{j}$ and $l_{k}$ is denoted as $\mathcal{E}_{(j,k)}$, and $\mathcal{E}_{(j,k)}=1$ if $l_{j}$ interferes with $l_{k}$; otherwise, $\mathcal{E}_{(j,k)}=0$.
Now, the objective is to find the set of links that can maximize the sum of the weights $w_{i}$ of each $l_{i}$.
This problem is a MWIS, and can be mathematically expressed as: maximizing $\sum_{\forall l_{i}\in\mathcal{L}}\nolimits w_{i}\mathcal{I}_{i}$ subject to $\mathcal{I}_{j} + \mathcal{I}_{k} \leq 1$, if $\mathcal{E}_{(j,k)}=1, \forall l_{j}\in\mathcal{L}, \forall l_{k}\in\mathcal{L}$
where $\mathcal{I}_{i}$ is a boolean index of $l_{i}$ which is $1$ if $l_{i}$ is scheduled (otherwise, $\mathcal{I}_{i}=0$). In our case, the weights are updated at each $t$, such that $w_{i} = Q_{i}(t) r_{i}(t)$, where $r_{i}(t)$ indicates the rate supported by link $\mathcal{l}_{i}$
at $t$. In fact, it is well-known that such max-weight policy achieves strong stability of the transmission queues, whenever the arrival rates are stabilizable, i.e., they fall inside the ergodic achievable rate region of the system~\cite{asilomar2012bethanabhotla}.
The exact value of $r_{i}(t)$ of D2D link $l_{i}$ cannot be obtained before a scheduling decision is made because the interference is unknown.
To circumvent this problem, $r_{i}(t)$ can be approximated as follows~\cite{jsac2011win}:
$r_{i}(t) = \log_{2}
\left(
    1+
    \frac{
        \mathcal{P}_{s_{i}\rightarrow d_{i}}(t)\left\|h_{i\rightarrow i}\right\|^{2}
    }{\sigma^{2} + \gamma}
\right)$
where
    $\mathcal{P}_{s_{i}\rightarrow d_{i}}(t)$ is the TX power from $s_{i}$ intended for $d_{i}$ at $t$,
    $h_{i\rightarrow i}$ is the (complex amplitude) channel gain from $s_{i}$ to $d_{i}$,
    $\sigma$ is the standard deviation of the (Gaussian) background noise,
    $\gamma$ is the interference thresholds, i.e., the maximum admissible interference $\gamma$ from a single interferer scheduled at the same time.
For solving MWIS problem, various approximation algorithms have been proposed because MWIS is NP-hard. Among them, we use well-established \textit{message-passing} that shows an excellent tradeoff between performance and complexity~\cite{tit2009sanghavi}.

\textbf{Distributed max-weight scheduling:}
In FlashLinQ, lower-priority D2D links can transmit only if they do not create significant interference to the higher-priority D2D links~\cite{ton2013wu}.
With the concept of max-weight scheduling, we set the priorities of D2D links as
$U_{i}(t) \triangleq \left\{r_{i}(t)\cdot Q_{i}(t)\right\}^{-1}$
which let each link wait $U_{i}(t)$ before transmission.

\begin{table}[t!]%
{\footnotesize
\caption{Expected Num. of Stalls ($\alpha=2$, $\gamma=5$\,dB)}
\label{tbl-prebuffering2}
    \centering %
	\begin{tabular}{r||r|r|r}
    \hline\hline
    PBT
    & mpMWIS-QP:
    & FlashLinQ-Q:
    & FlashLinQ-QP: \\
	\hline
     7 sec. &   0.7 &  2.4 &  0.8 \\
     8 sec. &   \textbf{0} [No Stalls] &  1.9 &  0.3 \\
     9 sec. &   \textbf{0} [No Stalls] &  1.1 &  \textbf{0} [No Stalls] \\
    10 sec. &   \textbf{0} [No Stalls] &  0.3 &  \textbf{0} [No Stalls] \\
	\hline\hline
	\end{tabular}
}
\end{table}

 \vspace{-2.5mm}
\subsection{Quality-Aware Streaming}\label{sec:streaming}

\textbf{Arrival process (placement of chunks):}
In each chunk time $\tau \in\{0,1,\cdots\}$, the TX of every link should place chunks within its queue. Different quality modes are available in each chunk, where a higher mode requires more bits.
We aim to maximize the total quality over all scheduled links subject to stability of the scheduled TX queues.
Let $\mathbb{P}(t) = \sum_{l_{i}\in\mathcal{L}}\mathbb{P}_{f_{i}}\left(q_{i}(t),t\right)$.
Then the following formulation is our objective function where (\ref{eq:meanratestable}) means all queues should fulfill rate stability~\cite{asilomar2012bethanabhotla}:
\begin{eqnarray}
\max & & \lim_{\tau\rightarrow \infty}\frac{1}{\tau}\sum_{t=0}^{\tau-1}\nolimits \mathbb{E}\left[\mathbb{P}(t)\right]\\
\text{subject to} & & \lim_{\tau\rightarrow \infty}\frac{1}{\tau}\sum_{t=0}^{\tau-1}\nolimits \mathbb{E}\left[Q_{i}(t)\right] < \infty, \forall l_{i} \in\mathcal{L}.\label{eq:meanratestable}
\end{eqnarray}

Now, the quality control involves choosing $q_{i}(t)$, the quality mode at chunk time $t$; as follows~\cite{asilomar2012bethanabhotla}:
\begin{equation}
\arg\max_{q_{i}(t)\in M}
\left[\mathbb{P}_{f_{i}}\left(q_{i}(t),t\right) -\alpha  \left\{\mathcal{N}\mathbb{B}_{f_{i}}(q_{i}(t), t)\right\} Q_{i}(t)\right].
\label{eq:final2}
\end{equation}

Thus,
$\lambda_{i}(t)$ can be $\mathcal{N}\mathbb{B}_{f_{i}}(q_{i}(t), t)$ (if $\tau \text{ mod } t = 0$, otherwise, $\lambda_{i}(t) = 0$) when the optimal $q$ is determined $\forall l_{i}$.

\textbf{Departure process (TX of bits):}
According to Shannon's capacity equation, $\mu_{i}(t)$ can be computed as:
$
\mu_{i}(t) = \mathcal{B}\cdot\log_{2}\left[1+\frac{\mathcal{P}_{s_{i}\rightarrow d_{i}}(t)\left\|h_{i\rightarrow i}\right\|^{2}}{\sigma^{2}+\sum_{j\neq i}
 \mathcal{P}_{s_{j}\rightarrow d_{i}}(t)\left\|h_{j\rightarrow i}\right\|^{2}}\right]
 \label{eq:dept}
$
where $\forall l_{i}, \forall l_{j} \in \mathcal{L}^{*}$ where $\mathcal{L}^{*}$ is the set of simultaneously scheduled links, $i\neq j$,
$\mathcal{P}_{s_{a}\rightarrow d_{b}}(t)$ is the power transmitted by $s_{a}$ intended for $d_{b}$, and $h_{j\rightarrow i}$ is the channel gain from the TX of $l_{j}$ to the RX of $l_{i}$ at $t$,
$\mathcal{B}$ is a bandwidth.
Depending on the distribution of used chunk sizes, the time scale of departure process is set independent to the time scale of arrival process.

\vspace{-1.0mm}
\section{Performance Evaluation}\label{sec:simulation}
In our simulations, we used 2-hour 4 video traces with realistic D2D layout and pathloss models~\cite{jsac2014ji}.
Our proposed centralized and distributed algorithms are denoted as mpMWIS-QP and FlashLinQ-QP.
They are compared with the FlashLinQ variant which controls the quality as in \textsection\ref{sec:streaming}, but the scheduling is randomized (named FlashLinQ-Q).

\textbf{Pre-Buffering Time (PBT):}
Table~\ref{tbl-prebuffering2} shows results when the PBT is from 7 to 10 (unit: sec).
We can see that mpMWIS-QP and FlashLinQ-QP present no stalls when PBTs=8 and PBT=9, i.e., mpMWIS-QP shows the best performance.
But, FlashLinQ-Q has stalls even if PBT=10.

\begin{figure}[t!]
	\begin{center}
		\includegraphics[scale=0.4]{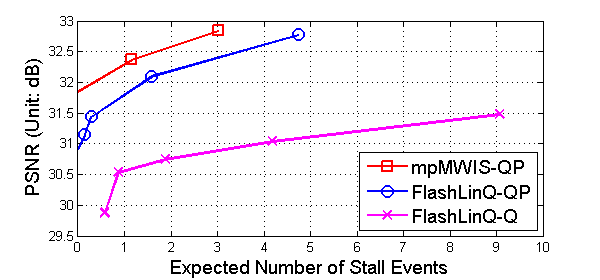}
	\end{center}
	\caption{Avg. Quality vs. Expected Num. of Stalls}
	\label{fig:stall2}
\end{figure}

\textbf{Average Quality vs. Num. of Stalls:}
Changing $\alpha$ allows to trade off quality with the number of stalls. For a given acceptable number of stalls,
Fig.~\ref{fig:stall2} shows mpMWIS-QP provides the highest PSNR; and the performance of FlashLinQ-QP is around $0.4$ dB lower than the one of mpMWIS-QP; FlashLinQ-Q shows around $1.6$ dB lower PSNR.

 \vspace{-1.0mm}
\section{Concluding Remarks}\label{sec:conclusion}
This paper proposed centralized quality-aware adaptive streaming and scheduling algorithms which can be used for D2D video delivery applications.
For basestation-aided centralized scheduling, a message-passing algorithm is used to obtain the solutions of the MWIS link scheduling problem..
For distributed scheduling, we improved a FlashLinQ scheduler with the principle of max-weight scheduling.
For streaming, a quality-aware stochastic algorithm is introduced.

 \vspace{-1.0mm}
\section*{Acknowledgement}\label{sec:conclusion}
Full extended version of this abstract is under review to be published in IEEE/ACM Transactions on Networking; and the title of the manuscript is ``Quality-Aware Streaming and Scheduling for Device-to-Device Video Delivery".

 \vspace{-1.0mm}


\begin{thebibliography}{1}
\bibitem{cm2013golrezaei} N. Golrezaei, A.F. Molisch, A.G. Dimakis, and G. Caire, ``Femtocaching and Device-to-Device Collaboration: A New Architecture for Wireless Video Distribution," \textit{IEEE Communications Magazine}, 51(4):142-149, April 2013.

\bibitem{ton2013wu} X. Wu, S. Tavildar, S. Shakkottai, T. Richardson, J. Li, R. Laroia, and A. Jovicic, ``FlashLinQ: A Synchronous Distributed Scheduler for Peer-to-Peer Ad Hoc Networks," \textit{IEEE/ACM Transactions on Networking}, 21(4):1215-1228, August 2013.

\bibitem{tit2009sanghavi} S. Sanghavi, D. Shah, and A.S. Willsky, ``Message Passing for Maximum Weight Independent Set," \textit{IEEE Transactions on Information Theory}, 55(11):4822-4834, November 2009.

\bibitem{asilomar2012bethanabhotla} D. Bethanabhotla, G. Caire, and M.J. Neely, ``Joint Transmission Scheduling and Congestion Control for Adaptive Streaming in Wireless Device-to-Device Networks" in \textit{Proc. of Asilomar Conference on Signals, Systems, and Computers}, CA, USA, November 2012.

\bibitem{jsac2014ji} M. Ji, G. Caire, and A.F. Molisch, ``Wireless Device-to-Device Caching Networks: Basic Principles and System Performance," Available on ArXiv: \url{http://arxiv.org/abs/1305.5216}, April 2014.

\bibitem{jsac2011win} M.Z. Win, P.C. Pinto, and L.A. Shepp, ``A Mathematical Theory of Network Interference and Its Applications," \textit{Proceedings of the IEEE}, 97(2):205-230, February 2009.


\end{thebibliography}
\end{document}